\documentclass[
  ,draft            
  ]
  {aipproc}

\layoutstyle{6x9}

\begin{document}

\title{The Subaru Coronagraphic Extreme AO Project}


\author{Frantz Martinache}{
  address={Subaru Telescope}
}

\author{Olivier Guyon}{
  address={Subaru Telescope}
}

\author{Julien Lozi}{
  address={Subaru Telescope}
}

\author{Vincent Garrel}{
  address={Subaru Telescope}
}

\author{C\'elia Blain}{
  address={Subaru Telescope}
}

\author{Gaetano Sivo}{
  address={Subaru Telescope}
}

\begin{abstract}
High contrast coronagraphic imaging is a challenging task for
telescopes with central obscurations and thick spider vanes, such as
the Subaru Telescope.
Our group is currently assembling an extreme AO bench designed as an
upgrade for the newly commissionned coronagraphic imager instrument
HiCIAO, that addresses these difficulties.
The so-called SCExAO system combines a high performance PIAA
coronagraph to a MEMS-based wavefront control system that will be used
in complement of the Subaru AO188 system.
We present and demonstrate good performance of two key optical
components that suppress the spider vanes, the central obscuration and
apodize the beam for high contrast coronagraphy, while preserving the
throughput and the angular resolution.
\end{abstract}

\maketitle

\section{Introduction}

In the first phase of the SCExAO project (spring 2010), a high
performance PIAA coronagraph will be implemented with a MEMs-based
wavefront control as an upgrade that will feed Subaru's newly
commissioned coronagraphic imager instrument HiCIAO
\citep{2008SPIE.7014E..42H} in the context of the Subaru Strategic
Exploration of Exoplanets and Disks (SEEDS) campain. SCExAO
\citep{2009arXiv0903.5001L} uses a unique combination of advanced
pupil remapping techniques: the Phase Induced Amplitude Apodization
(PIAA) Coronagraph \citep{2003A&A...404..379G, 2005ApJ...622..744G}
apodizes the pupil and removes the central obscuration of the beam,
and a Spider Removal Plate (SRP) almost eliminates the diffraction
spikes created by the spider vanes.
While the achieved raw contrast level ($\sim 10^5$) in an Extreme-AO
system is mostly driven by the speed and accuracy of the wavefront
control system, the coronagraphic approach described in this paper and
adopted for SCExAO will permit to achieve this raw contrast level at
angular separations down to 1 $\lambda/D$.

\section{Lossless Apodization}

On the Subaru Telescope pupil, the size of the central obstruction (30
\%, linear) and spider thickness (22cm) require that the coronagraph
is designed to remove the diffraction features they create in the
focal plane. The spiders, if left uncorrected, create 4 spikes at
$\sim 10^3$ contrast. These spikes are especially problematic at small
angular separation, where they cover most of the position angle space.
The central obscuration, at 30 \%, creates its own set of diffraction
rings with a $10^3$ contrast level at the peak of the central
obstruction's first ring (approximately 10 $\lambda/D$). Any
coronagraph designed to offer contrast better than $10^3$ on Subaru
Telescope therefore needs to take into account the spiders and
central obstruction.

\subsection{PIAA with central obscuration}

\begin{figure}
  \resizebox{.8\textwidth}{!}{\includegraphics{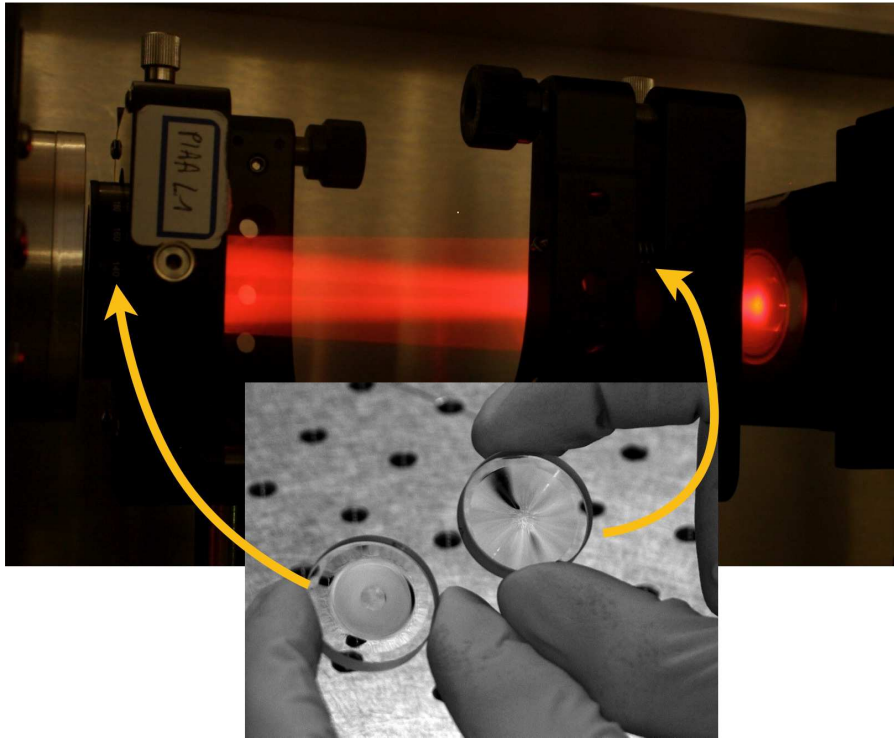}}
  \caption{
    Apodization of the beam by the PIAA lenses, a core component of
    the SCExAO bench. The lens L$_1$ (on the left), changes the
    distribution of light in the pupil plane and L$_2$ (on the right)
    compensates the distortions of the wavefront L$_1$
    introduces. Notice that as the light travels from left to right,
    the central obscuration of the telescope disappears and the pupil
    gets apodized, ready for high-performance coronagraphy.}
  \label{fig:piaalenses}
\end{figure}

The simplest way of performing an apodization is to insert a mask
whose radial transmission profile follows a prolate spheroidal
function, the so-called classical pupil apodization (CPA). 
While extremely robust, and insensitive to moderate tip-tilt
residuals, this approach has two main drawbacks: the throughput is low
(as low as $\sim$ 0.1 for a $10^{-10}$ contrast
\citep{2006ApJS..167...81G}), and the effective pupil diameter is
reduced by a factor approximately equal to the square root of the
throughput (due to the fact that apodizers remove the light mostly at
the edges of the pupil), which translates into a loss of angular
resolution.

The Phase Induced Amplitude Apodization (PIAA)
\citep{2003A&A...404..379G}
addresses these issues and apodizes the beam using a very different
approach. It uses a set of two tailored optics working in pair:
inserted in the pupil plane, the first (L$_1$ on
Fig. \ref{fig:piaalenses}) changes the distribution of light, while
the second (L$_2$) collimates the beam for an on-axis source.
The design retained for the SCExAO upgrade
(cf. Fig. \ref{fig:piaalenses}) suppresses the central obscuration.

Unlike CPA that discards light, the PIAA redistributes it across the
pupil and therefore preserves the throughput and the angular
resolution. Its main drawback is however the introduction of large
aberrations for off-axis sources, an issue we address toward the end
of the paper.

\subsection{Removing the spider vanes}

\begin{figure}
  \resizebox{.8\textwidth}{!}{\includegraphics{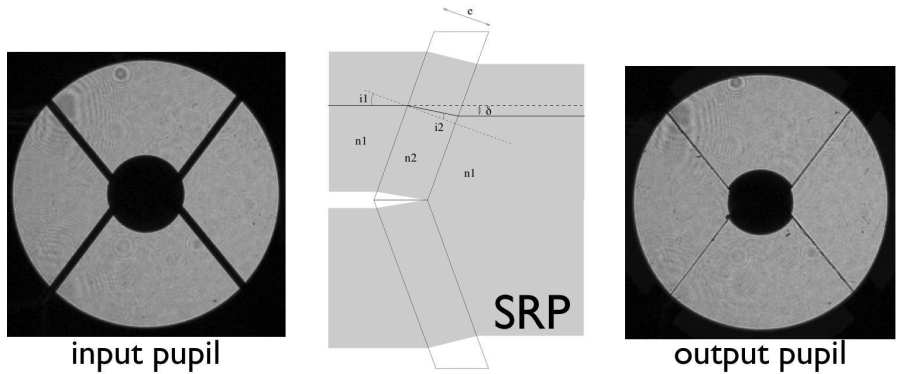}}
  \caption{Suppression of the spider vanes by the SRP. The four
    quadrant defined by the spider vanes of the Subaru Telescope pupil
  (left panel) are translated inward by the SRP, using straightforward
  geometric optics (central panel). In the output pupil (right panel),
  the spider vanes are considerably reduced so that the diffraction
  spikes they create won't affect the SCExAO contrast performance.}
  \label{fig:SRP}
\end{figure}

\begin{figure}
  \resizebox{.8\textwidth}{!}{\includegraphics{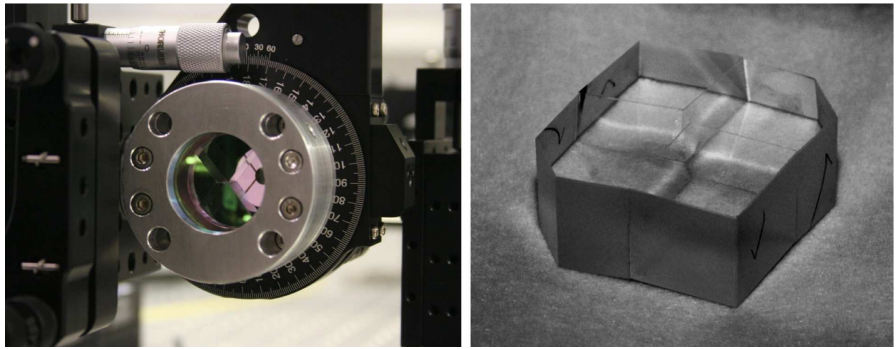}}
  \caption{Left-hand panel: SRP in its mount on the SCExAO bench;
    Right-hand panel: close-up of the assembled SRP. The plate is 15
    mm thick. Note that since the spider vane angle is not 45$^\circ$,
  the SRP is not continuous at the interface between the four plates.}
  \label{fig:SRPphoto}
\end{figure}

To suppress the diffraction by the spiders, our approach
(Fig. \ref{fig:SRP}), is to translate 
each of the four parts of the beam with a single tilted plate of
glass to fill the gap due to the spiders.
The spider vanes of the Subaru Telescope are 224 mm thick, for a total
pupil diameter of 7.92 m.
The SRP (Spider Removal Plate) consists of four tilted plane-parallel
plates, each translating a part of the pupil inwards, as shown in Figure
\ref{fig:SRP}. It can be best described as a ``pyramid-shaped
rooftop'' (cf. Fig. \ref{fig:SRPphoto}) of constant thickness.
All four plates were cut from the same plane-parallel plate
(a.k.a. optical window), to guarantee, within tolerances, a constant
thickness.

For a window of thickness $e = 15$ mm and index $n = 1.443$ (Fused
Silica for $\lambda = 1.6\mu$m), each plate needs to be tilted by an
angle $\alpha = 5.004 \pm 0.02^{\circ}$, to guarantee the continuity
of the wavefront on-axis after remapping within $\lambda/10$.

\subsection{Recovering the field of view}
\label{sec:inv}

\begin{figure}
  \resizebox{.75\textwidth}{!}{\includegraphics{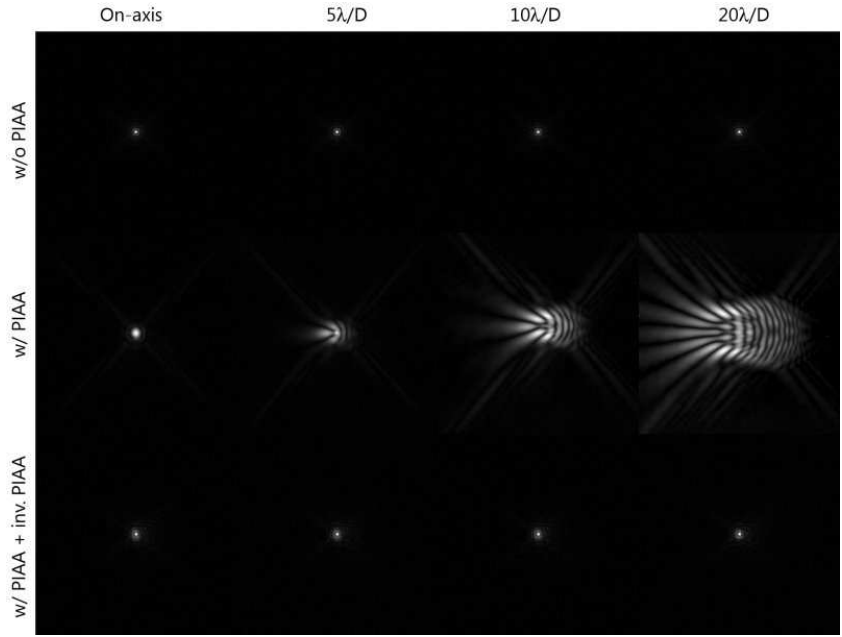}}
  \caption{Series of off-axis images taken with the SCExAO system in
    the lab. These images show that the inverse PIAA (bottom row)
    efficiently corrects the huge aberrations the PIAA introduces in
    the first place (middle row), and turns the pineapple-shaped
    off-axis images into more conventional Airy-like images, virtually
    identical to the ones the system produces with no beam apodization
    at all (top row).}
  \label{fig:fov}
\end{figure}

For an off-axis source, the pupil-remapping the PIAA performs
introduces large aberrations to the wavefront, which translate into
very unsually shaped PSFs. The matter is not new and has been
extensively described by \citet{2005ApJ...622..744G} in the case of a
non-obstructed aperture.
The novelty in the SCExAO PIAA is that it completely suppresses the
central obscuration. This however requires a somewhat brutal
remapping, with dramatic consequences on the PSF which finds itself
pineapple-shaped at separations greater than 10 $\lambda/D$
(cf. Fig. \ref{fig:fov}, middle row).
Fortunately, just like with any pupil remapping 
\citep{2003A&A...404..379G}, the aberrations the
PIAA introduces can entirely be corrected using an exact copy of the
PIAA, after the focal plane mask, only plugged backwards, which
restores the original pupil and provides wide field of view imaging
capabilities.
Fig. \ref{fig:fov} (bottom row) demonstrates the spectacular
efficiency of the inverse PIAA.

\bibliographystyle{aipproc}

\bibliography{martinache}

\end{document}